An infrared measurement of chemical desorption from interstellar ice analogues

Y. Oba[1]*, T. Tomaru[1], T. Lamberts[2], A. Kouchi[1] and N. Watanabe[1]

[1]Institute of Low Temperature Science, Hokkaido University, Sapporo, Hokkaido, Japan.

[2]Institute for Theoretical Chemistry, University of Stuttgart, Stuttgart, Germany.

*e-mail: oba@lowtem.hokudai.ac.jp

**In molecular clouds at temperatures as low as 10 K, all species except hydrogen and helium should be locked in the heterogeneous ice on dust grain surfaces. Nevertheless, astronomical observations have detected over 150 different species in the gas phase in these clouds. The mechanism by which molecules are released from the dust surface below thermal desorption temperatures to be detectable in the gas phase is crucial for understanding the chemical evolution in such cold clouds. Chemical desorption, caused by the excess energy of an exothermic reaction, was first proposed as a key molecular release mechanism almost 50 years ago[1]. Chemical desorption can, in principle, take place at any temperature, even below the thermal desorption temperature. Therefore, astrochemical net- work models commonly include this process[2,3]. Although there have been a few previous experimental efforts[4–6], no infrared measurement of the surface (which has a strong advantage to quantify chemical desorption) has been performed. Here, we report the first infrared in situ measurement of chemical desorption during the reactions H + H$_2$S → HS + H$_2$ (reaction 1) and HS + H → H$_2$S (reaction 2), which are key to interstellar sulphur chemistry[2,3]. The present study clearly demonstrates that chemical desorption is a more efficient process for releasing H$_2$S into the gas**

**phase than was previously believed. The obtained effective cross-section for chemical desorption indicates that the chemical desorption rate exceeds the photodesorption rate in typical interstellar environments.**

Astronomically abundant molecules, such as $H_2$, $H_2O$ and $CH_3OH$, can be synthesized on icy dust grains by hydrogen addition to H, O and CO, respectively[7–9], and some of these molecules are released into the gas phase from low-temperature grains. One of the non-thermal processes often proposed is photodesorption[7], which has been experimentally demonstrated to play an important role in photon-dominated regions[10–12]. However, photodesorption would be inefficient in dense regions where the ultraviolet field is very weak. Therefore, additional desorption processes need to be invoked. The importance of chemical desorption has been recognized mainly in the modelling community to explain the gas phase abundances of molecules that have originated on a surface. Indeed, chemical desorption with varying efficiencies has been incorporated into chemical models[2,3]. It is certainly the least well understood of the gas-grain processes that govern the results of astrochemical models to a large extent[13]. One might simply assume that most of the reaction products desorb from the surface after a reaction has taken place because, in general, the exothermicity (several eV) is significantly larger than the hydrogen bonding and physisorption energies (<0.5 eV) expected for adsorbates on dust grains. However, surprisingly, infrared in situ measurements for surface hydrogenation reactions; for example, CO hydrogenation to produce $H_2CO$ and $CH_3OH$, have indicated that most of these products remain on the surface[14,15]. Previous experiments focusing on quantifying the efficiency of chemical desorption were performed by gas phase detection solely with the use of a quadrupole mass spectrometer (QMS)[4–6]. In these experiments, the extent of desorption was derived only from the gas phase measurements. In contrast, infrared measurements can detect species on the substrate. Since chemical desorption is certainly a surface process, to monitor surface species by infrared measurements is crucial for a complete understanding of the chemical desorption phenomenon. In addition, time-resolved infrared measurements are necessary for obtaining chemical desorption cross-sections, which are highly desirable for chemical modelling studies. The present experiment, which takes an ideal reaction system where the reactant and product are both $H_2S$, enables us to quantify desorption after a reaction by in situ infrared absorption spectroscopy. In our reaction system, reaction 1 (R1) occurs between an $H_2S$ molecule and an H atom. R1 is expected to proceed by quantum tunnelling through a moderate activation barrier of about 1,560 K[16], while H atom addition to form $H_3S$ is unlikely to occur[17]. The reactions of HS with $H_2$ and $H_2O$ are both endothermic and are thus inhibited at low temperatures. As a result, HS on the ice can only react again with an H atom to produce $H_2S$ (reaction 2 (R2)). That is, we can eliminate the ambiguity arising from the uncertainties in the absorption coefficients of different parent and product molecules.

The experimental details are summarized in the Methods. Briefly, solid $H_2S$ is produced on an ice dust analogue, amorphous solid water (ASW), by gaseous $H_2S$ deposition at 10 K. This is subsequently exposed to H atoms, which were produced by the dissociation of $H_2$ with an efficiency of >60%. Variations in the $H_2S$ abundance are monitored in situ using a Fourier-transform infrared (FTIR) spectrometer in reflection mode. Furthermore, to confirm the FTIR measurements for chemical desorption, following atom exposure, a temperature programmed desorption (TPD) experiment is performed to measure both the unreacted parent $H_2S$ and the reaction product (R2) that are present on the surface. A ramping rate of 4 K min$^{-1}$ is used until a final temperature of 200 K is reached.

Figure 1 shows the FTIR spectrum of solid $H_2S$ (0.7 monolayer; 1 monolayer is equivalent to $1 \times 10^{15}$ molecules cm$^{-2}$) deposited on ASW at 10 K, as well as the difference spectra after exposure to H atoms for up to 120 min. Since the surface area of ASW under our experimental conditions is about ten times larger than that of a flat crystalline surface[15], the H2S coverage on ASW should correspond to ~0.07. Figure 1 clearly shows a decrease in the peak intensity for the S–H stretching band of solid $H_2S$ (2,570 cm$^{-1}$)[18,19] upon H atom exposure. No other sulphur-bearing species, such as $H_2S_2$ (~2,500 cm$^{-1}$) or $HS_2$ (2,483 cm$^{-1}$)[19] are observed besides $H_2S$. In addition, disulphur ($S_2$), which is infrared inactive, was not detected during the TPD experiment. Furthermore, we confirmed that no decrease in $H_2S$ is observed when solid $H_2S$ is exposed only to $H_2$ molecules on ASW. This indicates that the impact of $H_2$ on $H_2S$ induces neither a reaction nor desorption. The surface abundance of $H_2S$ gradually decreased, levelling off at 60% of the initial amount of $H_2S$ (Fig. 2), while 40% of the $H_2S$ remained on the surface after 120 min without desorbing. Figure 3 shows TPD spectra of $H_2S$ (m/z = 34) after exposure to H atoms or $H_2$ molecules for 120 min. Both spectra show three peaks at 95, 145 and 160 K, which are attributable to (1) the desorption of $H_2S$ from the top of the ASW surface, (2) $H_2S$ emitted via a molecular volcano process and (3) co-desorption with $H_2O$, respectively[20]. After 120 min, the total TPD peak area for H atom exposure was about 55% of that for $H_2$ exposure only.

Both the FTIR spectra and TPD measurements thus clearly demonstrate that chemical desorption is induced by reactions with H atoms. From the TPD measurement, however, the amount of remaining $H_2S$ is 15% larger than that measured by FTIR spectrometry because the TPD measurements detected $H_2S$ desorbed from not only the sample surface exposed to H atoms but also non-exposed areas of the cold head. The remnant undesorbed $H_2S$ may be related to the adsorption sites on the ASW surface, as previously reported[8,21]. Porous ASW is known to have various adsorption sites with different associated adsorption energies[8]. $H_2S$ molecules adsorbed in the potentially shallow sites may preferentially react with H atoms, leading to desorption. In contrast, the reaction of H atoms with $H_2S$ in the deeper sites may be significantly suppressed, resulting in the presence of unreacted $H_2S$ even after long exposure. This interpretation is supported by the fact that a larger fraction of $H_2S$ (~80%) desorbed from the surface of polycrystalline ice (Supplementary Fig. 1), which has fewer deep sites because of its surface flatness[8].

It is not a priori clear whether chemical desorption occurs during R1 or R2. The heats of reaction for R1 and R2 are ~0.6 and 3.8 eV, respectively. In principle, both reactions may cause desorption of products or intermediates because heats of reaction are larger than typical energies for physisorption and hydrogen bonding with the ASW surface. For the product of R1, the binding energy of HS to small water clusters is ~0.1 eV (see Methods), which is of the same order as the reaction heat. For R2, the binding energy of the $H_2S$ product to the same water clusters, as mentioned above, is also ~0.1 eV. When comparing differences in the heats of reaction and binding energies of R1 and R2, we deduce that desorption is more likely to be favourable in R2.

So far, we have assumed that both R1 and R2 occur. To confirm this, we performed additional experiments in which solid $H_2S$ was exposed to D atoms instead of H atoms on ASW at 10 K. The formation of deuterated sulphides (HDS and $D_2S$) was

then expected by successive H–D substitution reactions:

$$H_2S + D \rightarrow HS + HD \qquad (R3)$$

$$HS + D \rightarrow HDS \qquad (R4)$$

$$HDS + D \rightarrow DS + HD \qquad (R5)$$

$$DS + D \rightarrow D_2S \qquad (R6)$$

The infrared measurements and subsequent TPD experiments certainly confirm the formation of deuterated species (Fig. 4; see also Supplementary Fig. 2). This observation of HDS and/or $D_2S$ is clear evidence for the occurrence of not only R3–R6, but also R1 and R2. In this measurement, we did not quantify the chemical desorption because of the lack of experimental data for the absorption coefficient of HDS on ASW. In addition, we confirmed that when $D_2S$ was exposed to D atoms on the surface of amorphous solid $D_2O$ at 10 K, about 40% of the initial $D_2S$ was found to desorb after 120 min (Supplementary Fig. 3). The difference in the desorption fractions is not easy to explain simply by the mass difference between $H_2S$ and $D_2S$ and thus might originate in reaction dynamics.

In interstellar clouds, the main sulphur reservoir is commonly proposed to be present in an undetected form on dust grains[3,22,23]. Grain surface or ice chemistry is known to be of key importance in the interstellar formation of water and methanol and may also lead to more complex species[24]. $H_2S$ can be an important carrier of sulphur atoms in ice[25]. Modelling and observational studies also seem to indicate this[26,27]. So far, to the best of our knowledge, the models for $H_2S$ chemistry in the early stages of star formation either include chemical desorption with ~1% efficiency per reactive event or do not include this at all[2,3,23]. In the present experiment, we determine an effective chemical desorption cross-section of $2.2 \times 10^{-17}$ cm$^2$ by fitting the $H_2S$ attenuation to a single-exponential curve (Fig. 2; see Methods for the meaning of 'effective cross-section'). This value is comparable with the ultraviolet absorption cross-section of solid $H_2S$[28]. However, given that the photodesorption cross-section should be lower than the absorption cross-section, the effective chemical desorption cross-section may be larger than the photodesorption cross-section. In addition, assuming that the number density of H atoms is 10 cm$^{-3}$ at a temperature of 10 K in dense clouds, the H atom flux is estimated to be ~$10^5$ cm$^{-2}$ s$^{-1}$, which is about one to two orders of magnitude higher than the ultraviolet photon flux[29,30]. Under these conditions, the chemical desorption for $H_2S$ is likely to exceed photodesorption. As described above (see also Fig. 2), 40% of the initial deposited $H_2S$ remains on the surface while 60% of this is lost (desorbs) (see Methods for more details and the experimental meaning of 'efficiency'). When the chemical desorption efficiency, which includes desorption of both $H_2S$ and HS, is defined by the number of chemically desorbed molecules relative to the total number of initial molecules, its value is 60%.

We have successfully determined the effective cross-sections for the chemical desorption of $H_2S$ from ASW by in situ infrared measurements. We wish to stress, however, that the chemical desorption process depends on the specific reaction system at hand and, therefore, it will still be necessary to examine other molecules in the future. To fully understand surface reactions, it is essential in chemistry to know how exothermicity is used for desorption as well as dissipation into the bulk and/or in internal and kinetic motions of the molecule. Therefore, the present report has a significant impact on both astronomy and fundamental chemistry.

**Methods**

**Apparatus**. All experiments were performed using an ultra-high vacuum reaction system that was mainly composed of a stainless steel ultra-high vacuum chamber with a reflection-absorption-type FTIR spectrometer and a QMS. A gold-coated substrate was located at the centre of the apparatus, which was connected to a He cryostat. The substrate temperature could be controlled from 5 to 300 K. The base pressure inside the reaction chamber was of the order of $10^{-8}$ Pa. Gaseous species were deposited through a capillary plate at an angle of 45° to the surface normal.

**Experimental**. ASW was first produced on the substrate at 10 K by the vapour deposition of gaseous $H_2O$ at a deposition rate of ~2 monolayers $min^{-1}$, where one monolayer corresponds to $1 \times 10^{15}$ molecules $cm^{-2}$. The thickness of the ASW was adjusted to 30 monolayers based on the peak area and the band strength of the O–H stretching band of $H_2O$ at 3,300 $cm^{-1}$ (ref. 31). Gaseous $H_2S$ was then deposited on the ASW at 10 K at a deposition rate of 1 monolayer $min^{-1}$. The thickness of solid $H_2S$ was adjusted to 0.7 monolayer, which was evaluated based on the peak area and the band strength of the S–H stretching band at ~2,570 $cm^{-1}$ (ref. 18). The formed $H_2S$ was exposed to H atoms, which were produced by the dissociation of $H_2$ molecules in a microwave-induced plasma. The dissociation fraction of $H_2$ in the plasma was more than 60% under these experimental conditions[32]. The H atoms were cooled to ~100 K before leaching to the substrate through multiple collisions with the inner wall of an aluminium pipe, which was cooled to the liquid $N_2$ temperature. As a blank experiment, solid $H_2S$ was exposed to $H_2$ molecules only for 120 min to evaluate the effect of the co-deposited $H_2$ on the desorption of $H_2S$. This experiment is denoted as a H blank experiment.

An additional experiment was performed on polycrystalline $H_2O$ (c-$H_2O$) ice, which was prepared by the production of ASW at 10 K with a thickness of 30 monolayers, followed by increasing the substrate temperature to 150 K and subsequently reducing it back to 10 K. Next, 0.7 monolayer $H_2S$ was deposited on the c-$H_2O$ and this was exposed to H atoms for up to 120 min. Solid $H_2S$ (0.7 monolayer) was exposed to D atoms to confirm whether reactions R3–R6 occur on ASW (30 monolayers). D atoms were produced by a similar method to that for H atoms, where $D_2$ molecules were dissociated in microwave-induced plasma and cooled to 100 K.

Chemical desorption of $D_2S$ was also studied in another experiment. Gaseous $D_2S$ was deposited on pre-deposited deuterated water ($D_2O$) at 10 K. The abundances of $D_2O$ and $D_2S$ were adjusted to those of $H_2O$ and $H_2S$ in the former experiments based on the peak area of the O–D (2,570 $cm^{-1}$) and S–D stretching bands (1,860 $cm^{-1}$) and their band strengths, respectively[18,33]. The prepared solid sample was exposed to cold D atoms for up to 120 min. As a blank experiment, solid $D_2S$ was exposed to $D_2$ molecules only, for which we confirmed no desorption of $D_2S$ even after $D_2$ exposure for 120 min. The fluxes of H and D atoms were estimated as $1.3 \times 10^{14}$ and $2.6 \times 10^{14}$ atoms $cm^{-2}\,s^{-1}$, respectively, according to a procedure described previously[34].

After the atom or molecule exposure experiments, we performed a TPD experiment, in which each sample was heated to 200 K at a constant ramping rate of 4 K $min^{-1}$ and the desorbed species from the substrate were recorded with the QMS.

**Efficiency of chemical desorption**. There can be several definitions for the chemical desorption efficiency. In previous studies where a QMS was used as a tool for estimating chemical desorption, the chemical desorption efficiencies were generally derived from the fraction of the target molecule remaining after reactions on the substrate[4,5,35,36]. For comparison with previous studies, here, we use a similar definition; that is, the number of chemically desorbed molecules relative to the total number of initial molecules at the termination of reactions. Due to some potential difficulties on QMS, the derived chemical desorption efficiencies often have large errors[5]. In contrast, owing to the visibility of the $H_2S$ decrease on the infrared spectrum (Fig. 1), the desorption efficiency in the present study was—regardless of desorbed species—determined to be exactly 60% under the present experimental conditions (Fig. 2). However, it should be noted that the value can change with the fluence of atoms.

There is difficulty when modellers take the value of efficiency from experimental works because the definition of 'per reactive event' is generally used in the models while experiments on the surface reaction never provide values per reactive event. Unlike for the gas-phase beam experiment, the single collision condition cannot be controlled in experiments on surface reactions. Instead, we propose the additional chemical desorption efficiency under a different definition: a desorption yield per incident H atoms. In the earlier stages of the reaction (<5 min), the column density of $H_2S$ linearly decreased with H atom exposure time (Fig. 2). During this period, the decrease of $H_2S$ via chemical desorption was ~$2.1 \times 10^{14}$ molecules cm$^{-2}$. Since the H atom fluence during 5 min is $3.9 \times 10^{16}$ atoms cm$^{-2}$, the chemical desorption yield per incident H atom was calculated to be $5.4 \times 10^{-3}$. However, since most of the H atoms land on ASW and should recombine to yield $H_2$ when the flux of H atoms is very high[37], the actual number of H atoms that reacted with $H_2S$ should be much smaller than the calculated H fluence (that is, $3.9 \times 10^{16}$ atoms cm$^{-2}$). Therefore, the above calculated efficiency should be the lower limit under the present experimental conditions.

Moreover, in realistic interstellar environments where the H atom flux is much lower than under laboratory conditions, each H atom landing on a dust grain has a greater chance of encountering reactants besides other H atoms. As a result, the chemical desorption efficiency defined by per incident H atom should be sufficiently enhanced under realistic conditions. However, note again that the chemical desorption yield per incident H atom is not equivalent to that per reactive event.

**Effective cross-section**. By simply assuming that the chemical desorption of $H_2S$ proceeds through a single process such as photodesorption, the relative abundance of $H_2S$ can be represented as a single exponential decay curve as follows:

$$\Delta[H_2S]_t / [H_2S]_0 = A(1-\exp(-\sigma\varphi t)) \qquad (1)$$

where $\Delta[H_2S]_t$, $A$, $\sigma$ and $\phi$ represent the variation in the abundance of $H_2S$ at time $t$, a saturation value for the desorption fraction of $H_2S$, the effective cross-section of chemical desorption in $cm^2$ and the flux of H atoms ($\sim 1.3 \times 10^{14}$ atoms $cm^{-2}$ $s^{-1}$), respectively. By fitting the plots in Fig. 2 into equation (1), we obtain $\sigma = 2.2 \times 10^{-17}$ $cm^2$ for the chemical desorption of $H_2S$ from ASW, as described in the main text. Likewise, the effective cross-sections for the chemical desorption of $H_2S$ from c-$H_2O$ and $D_2S$ from amorphous solid $D_2O$ were calculated to be $5.3 \times 10^{-18}$ and $3.5 \times 10^{-18}$ $cm^2$, respectively. This assumption may be optimistic; nevertheless, we believe that the obtained order of magnitude of the effective cross-section is sufficient to conclude that chemical desorption of $H_2S$ would be very effective even in dense clouds where the H atom number density is much lower than that in the experiment, and thus the ratio of the collision frequency between H atoms and $H_2S$ to that between two H atoms, frequency(H–$H_2S$)/frequency(H–H), is enhanced.

**Binding energy of HS and H₂S**. To estimate the binding energies of HS and H$_2$S on a water surface, we optimized two clusters consisting of seven water molecules, both with and without the adsorbates. The binding energy is defined as the energy difference between the gas phase adsorbate separated from the relaxed water cluster and the total energy of the relaxed adsorbate-cluster system. The MPWB1K functional[38] and def2-TZVP basis set[39] were used in the implementation of Gaussian 09 (ref. 40) based on the benchmark performed by Lamberts and Kästner[16]. All density functional theory calculations comprise full geometry optimizations and all stationary points were verified by Hessian calculations; that is, by having zero imaginary frequencies.

In Supplementary Table 1, the structures of the adsorbate-cluster systems for the two water clusters and both adsorbates are displayed with their corresponding binding energies and zero-point energy corrections. The total binding energies are all of the same order of magnitude, ~10 kJ mol$^{-1}$; that is, ~0.1 eV. The interaction energy between a dimer of H$_2$S and H$_2$O is roughly 6 kJ mol$^{-1}$ (ref. 16). Note that the binding energy also contains a contribution related to the restructuring of the water cluster.

**References**

1. Williams, D. A. Physical adsorption processes on interstellar graphite grains. Astrophys. J. 151, 935–943 (1968).

2. Garrod, R. T., Wakelam, V. & Herbst, E. Non-thermal desorption from interstellar dust grains via exothermic surface reactions. Astron. Astrophys. 467, 1103–1115 (2007).

3. Vidal, T. H. G. et al. On the reservoir of sulphur in dark clouds: chemistry and

elemental abundance reconciled. Mon. Not. R. Astron. Soc. 489, 435–447 (2017).

4.  Dulieu, F. et al. How micron-sized dust particles determine the chemistry of our Universe. Sci. Rep. 3, 1338 (2013).

5.  Minissale, M., Dulieu, F., Cazaux, S. & Hocuk, S. Dust as interstellar catalyst. Quantifying the chemical desorption process. Astron. Astrophys. 585, A24 (2016).

6.  He, J., Emtiza, S. M. & Vidali, G. Mechanism of atomic hydrogen addition reactions on np-ASW. Astrophys. J. 851, 104 (2017).

7.  Tielens, A. G. G. M. The molecular universe. Rev. Mod. Phys. 85, 1021–1081 (2013).

8.  Hama, T. & Watanabe, N. Surface processes on interstellar amorphous solid water: adsorption, diffusion, tunneling reactions, and nuclear-spin conversion. Chem. Rev. 113, 8783–8839 (2013).

9.  van Dishoeck, E. F., Herbst, E. & Neufeld, D. A. Interstellar water chemistry: from laboratory to observations. Chem. Rev. 113, 9043–9085 (2013).

10. Hama, T. et al. A desorption mechanism of water following vacuum- ultraviolet irradiation on amorphous solid water at 90 K. J. Chem. Phys. 132, 164508 (2010).

11. Fayolle, E. C. et al. CO ice photodesorption: a wavelength-dependent study. Astrophys. J. Lett. 739, L36 (2011).

12. Muñoz Caro, G. M. et al. New results on thermal and photodesorption of CO ice using the novel InterStellar Astrochemistry Chamber (ISAC). Astron. Astrophys. 522, A108 (2010).

13. Cuppen, H. H. et al. Grain surface models and data for astrochemistry. Space Sci. Rev. 212, 1–58 (2017).

14. Watanabe, N., Nagaoka, A., Shiraki, T. & Kouchi, A. Hydrogenation of CO on

pure solid CO and CO–$H_2O$ mixed ice. Astrophys. J. 616, 638–642 (2004).

15. Hidaka, H., Miyauchi, N., Kouchi, A. & Watanabe, N. Structural effects of ice grain surfaces on the hydrogenation of CO at low temperatures. Chem. Phys. Lett. 456, 36–40 (2008).

16. Lamberts, T. & Kästner, J. Tunneling reaction kinetics for the hydrogen abstraction reaction H + $H_2S$ → $H_2$ + HS in the interstellar medium. J. Phys. Chem. A 121, 9736–9741 (2017).

17. Qi, J., Lu, D., Song, H., Li, J. & Yang, M. Quantum and quasiclassical dynamics of the multi-channel H + H2S reaction. J. Chem. Phys. 146, 124303 (2017).

18. Fathe, K., Holt, J. S., Oxley, S. P. & Pursell, C. J. Infrared spectroscopy of solid hydrogen sulfide and deuterium sulfide. J. Phys. Chem. A 110, 10793–10798 (2006).

19. Jiménez-Escobar, A. & Muñoz Caro, G. M. Sulfur depletion in dense clouds and circumstellar regions. I. $H_2S$ ice abundance and UV-photochemical reactions in the $H_2O$-matrix. Astron. Astrophys. 536, A91 (2011).

20. Collings, M. P. et al. A laboratory survey of the thermal desorption of astrophysically relevant molecules. Mon. Not. R. Astron. Soc. 354, 1133–1140 (2004).

21. Watanabe, N. et al. Direct measurements of hydrogen atom diffusion and the spin temperature of nascent $H_2$ molecule on amorphous solid water. Astrophys. J. Lett. 714, L233–L237 (2010).

22. van der Tak, F. F. S., Boonmanm, A. M. S., Braakman, R. & van Dishoeck, E. F. Sulphur chemistry in the envelopes of massive young stars. Astron. Astrophys. 412, 133–145 (2003).

23. Woods, P. M. et al. A new study of an old sink of sulphur in hot molecular cores: the sulphur residue. Mon. Not. R. Astron. Soc. 450, 1256–1267 (2015).

24. Taquet, V., Charnley, S. B. & Sipilä, O. Multilayer formation and evaporatin of deuterated ices in prestellar and protostellar cores. Astrophys. J. 791, 1 (2014).

25. Calmonte, U. et al. Sulphur-bearing species in the coma of comet 67P/ Churyumov–Gerasimenko. Mon. Not. R. Astron. Soc. 462, S253–S273 (2016).

26. Esplugues, G. B., Viti, S., Goicoechea, J. R. & Cernicharo, J. Modelling the sulphur chemistry evolution in Orion KL. Astron. Astrophys. 567, A95 (2014).

27. Holdship, J. et al. H2S in the L1157-B1 bow shock. Mon. Not. R. Astron. Soc. 463, 802–810 (2016).

28. Cruz-Diaz, G. A., Muñoz Caro, G. M., Chen, Y.-J. & Yih, T.-S. Vacuum-UV spectroscopy of interstellar ice analogs I. Absorption cross-sections of polar-ice molecules. Astron. Astrophys. 562, A119 (2014).

29. Prasad, S. S. & Tarafdar, S. P. UV radiation field inside dense clouds: its possible existence and chemical implications. Astrophys. J. 267, 603–609 (1983).

30. Shen, C. J., Greenberg, J. M., Schutte, W. A. & van Dishoeck, E. F. Cosmic ray induced explosive chemical desorption in dense clouds. Astron. Astrophys. 415, 203–215 (2004).

31. Gerakines, P. A., Schutte, W. A., Greenberg, J. M. & van Dishoeck, E. F. The infrared band strengths of $H_2O$, $CO$ and $CO_2$ in laboratory simulations of astrophysical ice mixtures. Astron. Astrophys. 296, 810–818 (1995).

32. Hama, T. et al. The mechanism of surface diffusion of H and D atoms on amorphous solid water: existence of various potential sites. Astrophys. J. 757, 185 (2012).

33. Miyauchi, N. et al. Formation of hydrogen peroxide and water from the reaction of cold hydrogen atoms with solid oxygen at 10 K. Chem. Phys. Lett. 456, 27–30 (2008).

34. Oba, Y., Osaka, K., Watanabe, N., Chigai, T. & Kouchi, A. Reaction kinetics and isotope effect of water formation by the surface reaction of solid $H_2O_2$ with H atoms at low temperatures. Faraday Discuss. 168, 185–204 (2014).

35. Minissale, M. & Dulieu, F. Influence of surface coverage on the chemical desorption process. J. Chem. Phys. 141, 014304 (2014).

36. Minissale, M., Moudens, A., Baouche, S., Chaabouni, H. & Dulieu, F. Hydrogenation of CO-bearing species on grains: unexpected chemical desorption of CO. Mon. Not. R. Astron. Soc. 458, 2953–2961 (2016).

37. Watanabe, N. & Kouchi, A. Ice surface reactions: a key to chemical evolution in space. Prog. Surf. Sci. 83, 439–489 (2008).

38. Zhao, Y. & Truhlar, D. G. Hybrid meta density functional theory methods for thermochemistry, thermochemical kinetics, and noncovalent interactions: the MPW1B95 and MPWB1K models and comparative assessments for hydrogen bonding and van der Waals interactions. J. Chem. Phys. A 108, 6908–6918 (2004).

39. Weigend, F., Häser, M., Patzelt, H. & Ahlrichs, R. RI-MP2: optimized auxiliary basis sets and demonstration of efficiency. Chem. Phys. Lett. 294, 143–152 (1998).

40. Frisch, M. J. et al. Gaussian 09 Revision D.01 (Gaussian, Wallingford, CT, 2016).

Acknowledgements

The authors thank H. Hidaka, T. Hama and J. Kästner for discussions about chemical desorption. This work was partly supported by a Japan Society for the Promotion of

Science Grant-in-Aid for Specially Promoted Research (JP17H06087) and Grant-in-Aid for Young Scientists (A) (JP26707030). Computational resources were provided by the state of Baden-Württemberg through bwHPC and the German Research Foundation through grant number INST 40/467-1 FUGG.

## Author contributions

Y.O. planned the experiments in consultation with N.W. and A.K. Y.O. and T.T. performed the experiments. T.L. performed the computational calculations on binding energy. All authors discussed the results. Y.O., T.L. and N.W. wrote the paper.

## Additional information

Supplementary information accompanies this paper at https://doi.org/10.1038/s41550-018-0380-9.

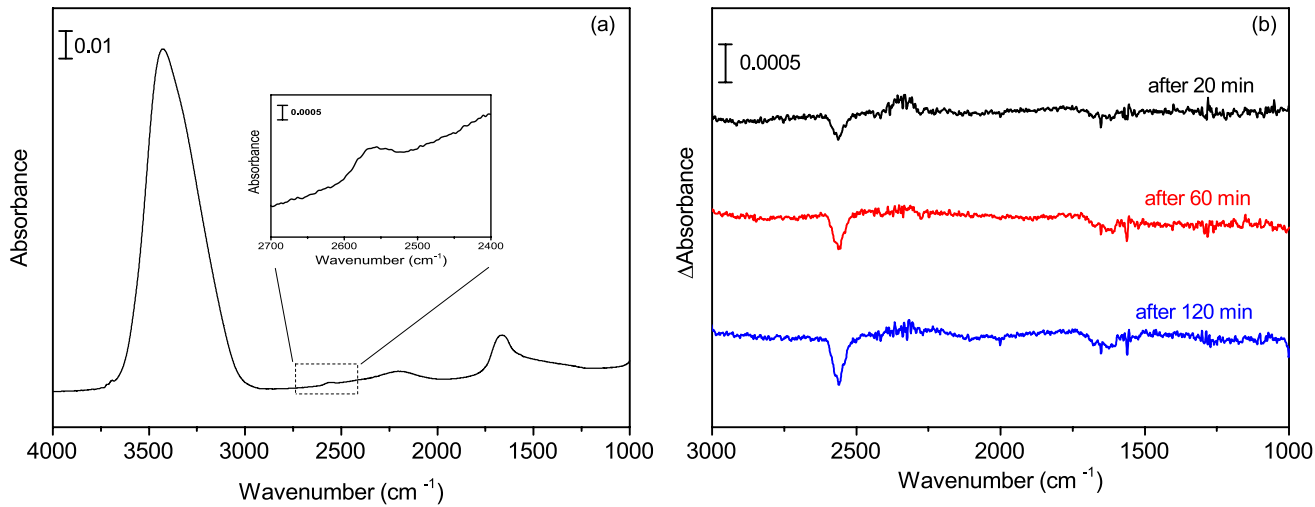

**Fig. 1 FTIR spectra of samples. a**. Solid $H_2S$ (=0.7 monolayer) deposited on ASW (~30 monolayers). The inset shows an enlarged spectrom focusing on the S-H stretching band of $H_2S$ (2,570 cm$^{-1}$)[18]. Difference spectra after exposure to H atoms for 20 (top), 60 (middle) and 120 min (bottom). **b**. The spectral change at around 2,300 cm$^{-1}$ is due to the temperal variation of atmospheric $CO_2$ concentration in the infrared beam line outside the vacuum chamber.

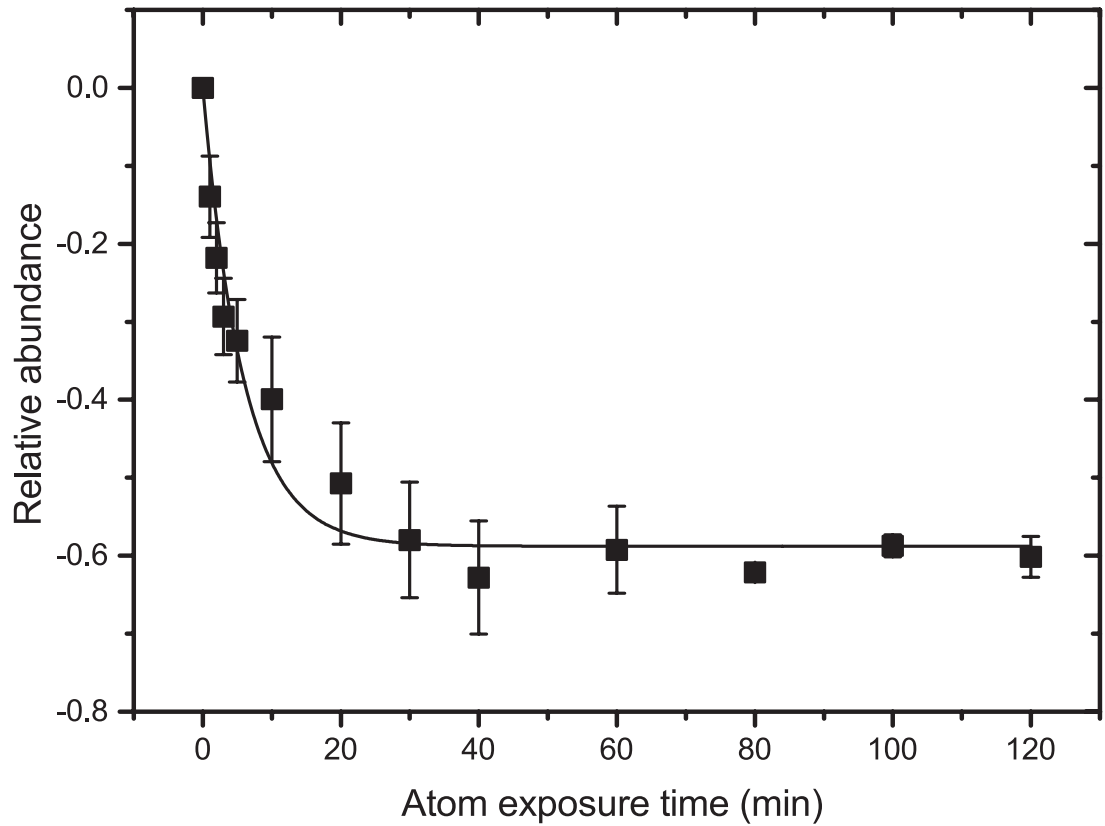

**Fig. 2 Variations in the relative abundance of solid $H_2S$ on ASW with relevance to H atom exposure times.** The solid line is a single exponential fit to equation (1) (see MEthods). Each data point is an average of three measurements, with the s.d. of the mean values indicated by the error bars.

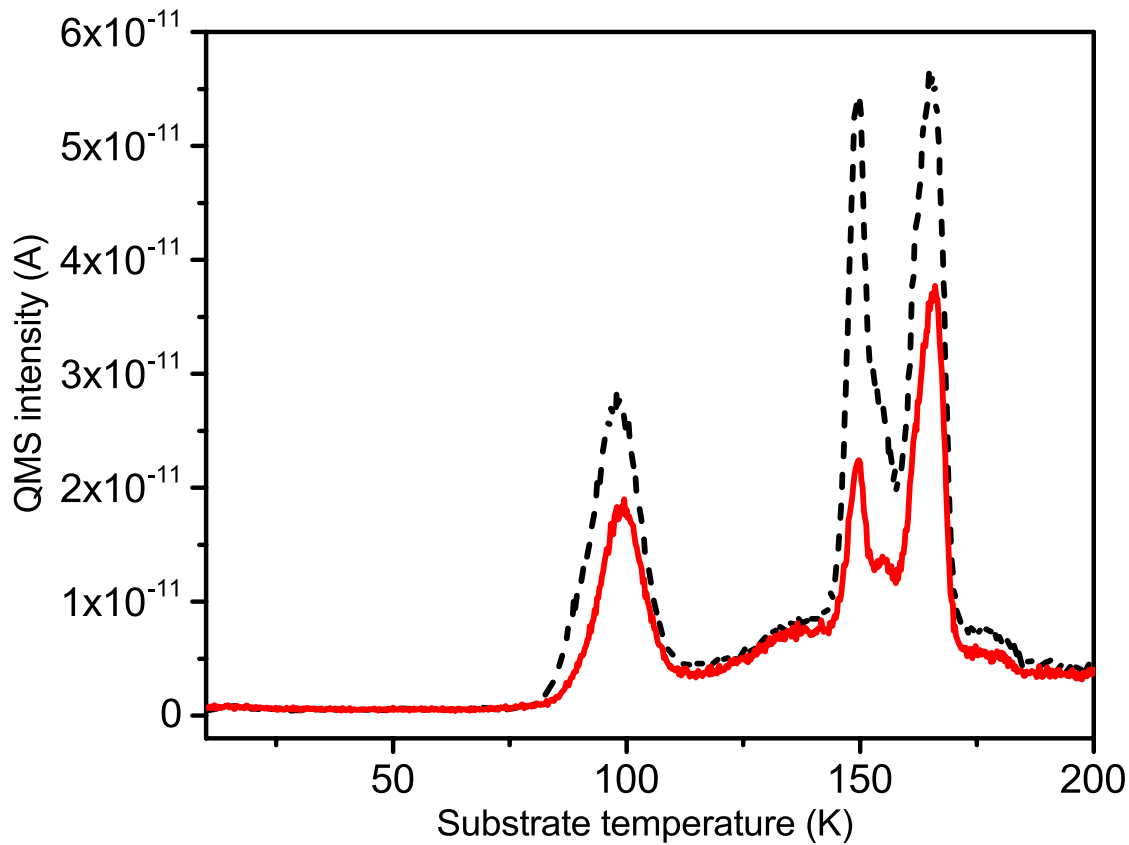

**Fig. 3 TPD spectra**. Solid H$_2$S (m/z = 34) from ASW obtained after exposure to H atoms (solid red line) and H2 molecules without H atoms (H-blank experiment; dashed black line).

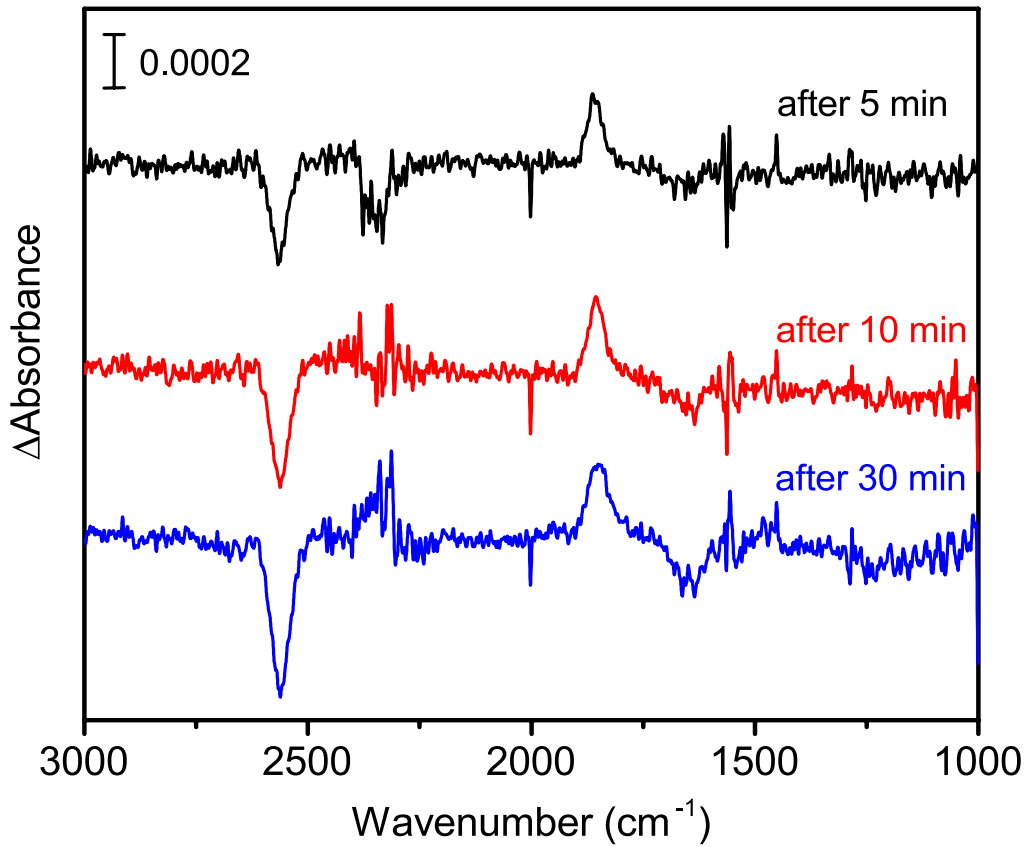

**Fig. 4 Difference spectra of solid H₂S after exposure to D atoms for varying lengths of time on ASW.** Top: 5 min. Middle: 10 min. Bottom: 30 min. The S-D stretching band of deuterated sulphi8des was observed at ~!,860 cm⁻¹. The spectral change at around 2,300 cm⁻¹ is due to the temporal variation of the atmospheric CO₂ concentration in the infrared beam line outside the cacuum chamber.